\documentclass{ws-procs9x6}

\begin{document}

\title{Exceptional and non-exceptional contributions to the radiative $\pi$ decay\footnote{Talk given at the 5th International Workshop on Chiral Dynamics, Theory and Experiment, 18-22 September (2006), Durham/Chapel Hill, NC, (USA);
IFIC/06$-$43 FTUV/06$-$1121 report. To appear in the Proceedings.}
}

\author{V.~Mateu}

\address{Departament de F\'\i sica Te\`orica-IFC, Universitat de Val\`encia-CSIC,\\
$^*$E-mail: mateu@ific.uv.es\\
}

\begin{abstract}
We have studied the spin-one resonance dominated form factors governing the radiative decay of the $\pi$, within the framework of resonance chiral theory. We obtain predictions for their value at zero momentum transfer and also a description of their $q^2$ dependence.
We also fit the possible new physics tensor coupling. 
\end{abstract}


\bodymatter

\section{Main results and conclusions}
\noindent
For many years the radiative $\pi$ decay (RPD) process ($\pi\to e \nu \gamma$) has been an open window for the speculation about new physics (NP) \cite{Belyaev:1991gs}. 
The available experimental data \cite{ISTRA-PIBETA} seemed to indicate a deficit of events in one 
region 
when compared to the Standard Model (SM) prediction.

The physical description of the process can be split into two pieces: the inner bremsstrahlung (IB) contribution contribution 
and the structure-dependent (SD) contribution, described by two form factors. 

Still we lack a complete analysis of the process including $q^2$ dependence in the form factors, and this is the main purpose of this work. We will use the resonance chiral theory (R$\chi$T) for the description of the form factors \cite{Mateu-Portoles}.

Within the SM the SD contributions can be parametrized by a vector form factor and a axial-vector form factor. A tensor form factor would interfere destructively in the region in which the deficit of events occurs.

The axial-vector form factor has already been calculated \cite{Cirigliano:2004ue}. 
The VVP Green Function 
so far has been calculated only with one multiplet of vector meson resonances and it is not able to fulfil at the same time all the short distance constraints
\cite{VVP}. Including a second multiplet we can satisfy all the requirements and we obtain a reliable vector form factor \cite{Mateu-Portoles}:
\begin{equation}
F_V(q^2) \, = \, - \, \frac{m_{\pi^+}}{3 \, \sqrt{2} \, B_0 \, F \, 
M_{V_1}^2 \, M_{V_2}^2} \, \frac{c_{000} \, + \, 
c_{010} \, q^2}{(M_{V_1}^2 - q^2)\,(M_{V_2}^2 - q^2)} \;
\end{equation} 
were $c_{000}$ is fixed by the anomaly and $c_{010}$ is fitted from the $\pi\to\gamma\gamma^*$ phenomenology.

Finally with this framework we calculate the tensor hadronic matrix element
, usually expressed through the so called magnetic susceptibility $\chi$ and the  vector meson couplings to vector and tensor currents:
\begin{equation}
\chi=-\dfrac{2}{M_V^2}\backsimeq-3.37\,\mathrm{GeV}^{-2},\quad \dfrac{f^\perp_{V}}{f_{V}}=\dfrac{\left\langle \bar{q}q\right\rangle }{M_V F_V^2}\backsimeq 0.78(1).
\end{equation}
With the new but preliminary data from the PIBETA collaboration \cite{dinko} the mismatch now is gone, as can be seen in tab.~\ref{tabladatos}.
\begin{table}
\caption{Comparison of the theoretical predictions and the experimental data }
\begin{tabular}[tbh]{|c|c|c|c|c|}
\hline $E_{e^+}^{\mathrm{min}}$(MeV) & $E_{\gamma}^{\mathrm{min}}$(MeV) & 
$\theta_{e\gamma}^{\mathrm{min}}$ & $R_{\mathrm{exp}}(\times 10^{-8})$\cite{dinko}& $R_{\mathrm{the}}$\cite{Mateu-Portoles}\\ 
\hline 50 & 50 & -- & $2.655\pm0.058$& 2.55(9)\\ 
\hline 10 & 50 & $40\,{}^\circ$ & $14.59\pm0.60$&14.66(1)\\ 
\hline 50 & 10 & $40\,{}^\circ$ & $37.95\pm 0.28$& 37.70(98)\\ 
\hline 
\end{tabular}\label{tabladatos}
\end{table}
Nevertheless we fit the value for the NP tensor coupling and we get:
\begin{equation}
f_T=(4\pm 6)\times 10^{-4}
\end{equation} 
which is compatible with zero and with the value dictated by SUSY\cite{Belyaev:1991gs}.
\section*{Acknowledgements}\noindent
Work supported by
EU FLAVIAnet  (MRTN-CT-2006-035482)
, the Spanish MEC
(FPA2004-00996), Generalitat Valenciana
(GRUPOS03/013 and GV05-164) and by ERDF funds from the EU Commission.

\end{document}